\documentclass[reprint,amsmath,amssymb,aps,prl,superscriptaddress,showpacs]{revtex4-1}

\usepackage{amsfonts}
\usepackage{amsmath}
\usepackage{amssymb}
\usepackage{epsf, epsfig, graphics}
\usepackage{graphicx}
\usepackage{subfigure}
\usepackage{mathtools}
\usepackage{color}
\usepackage{dcolumn}
\usepackage{bm}
\usepackage{ulem}
\graphicspath{{./Figures/}}

\usepackage{hyperref}
\makeatletter
\def\UrlAlphabet{%
      \do\a\do\b\do\c\do\d\do\e\do\f\do\g\do\h\do\i\do\j%
      \do\k\do\l\do\m\do\n\do\o\do\p\do\q\do\r\do\s\do\t%
      \do\u\do\v\do\w\do\x\do\y\do\z\do\A\do\B\do\C\do\D%
      \do\E\do\F\do\G\do\H\do\I\do\J\do\K\do\L\do\M\do\N%
      \do\O\do\P\do\Q\do\R\do\S\do\T\do\U\do\V\do\W\do\X%
      \do\Y\do\Z}
\def\UrlDigits{\do\1\do\2\do\3\do\4\do\5\do\6\do\7\do\8\do\9\do\0}
\g@addto@macro{\UrlBreaks}{\UrlOrds}
\g@addto@macro{\UrlBreaks}{\UrlAlphabet}
\g@addto@macro{\UrlBreaks}{\UrlDigits}
\makeatother

\begin{document}

\title{Glass stability changes the nature of yielding under oscillatory shear}

\author{Wei-Ting~Yeh}
\affiliation{Department of Physics, Nagoya University, 464-8602 Nagoya, Japan}

\author{Misaki Ozawa}

\affiliation{Laboratoire Charles Coulomb (L2C), Universit\'e de Montpellier, CNRS, 34095 Montpellier, France}

\affiliation{Laboratoire de Physique de l'Ecole Normale Sup\'erieure, ENS, Universit\'e PSL, CNRS, Sorbonne Universit\'e, Universit\'e de Paris, F-75005 Paris}

\author{Kunimasa Miyazaki}
\affiliation{Department of Physics, Nagoya University, 464-8602 Nagoya, Japan}

\author{Takeshi Kawasaki}
\affiliation{Department of Physics, Nagoya University, 464-8602 Nagoya, Japan}

\author{Ludovic Berthier}

\affiliation{Laboratoire Charles Coulomb (L2C), Universit\'e de Montpellier, CNRS, 34095 Montpellier, France}

\affiliation{Department of Chemistry, University of Cambridge, Lensfield Road, Cambridge CB2 1EW, United Kingdom}

\date{\today}


\begin{abstract}
We perform molecular dynamics simulations to investigate the effect of
a glass preparation on its yielding transition under oscillatory shear. We use swap Monte Carlo to investigate a broad range of glass stabilities from poorly annealed to highly stable systems. We observe a qualitative change in the nature of yielding, which evolves from ductile to brittle as glass stability increases. Our results disentangle the relative role of mechanical and thermal annealing on the mechanical properties of amorphous solids, which is relevant for various experimental situations from the rheology of soft materials to fatigue failure in metallic glasses.
\end{abstract}

\maketitle


Amorphous solids are generated by decreasing the temperature or by
increasing the pressure of supercooled liquids, colloidal suspensions,
and granular materials~\cite{Debenedetti_2001, O'Hern_2002,
Liu_1998}. 
When deformed, their mechanical response is initially elastic 
but plastic deformations appear at larger strains, which eventually trigger material flow. 
This elastic-to-plastic transformation is a yielding transition, which is currently an active research topic~\cite{Bonn_2017,Nicolas_2018}. 
Understanding the microscopic nature of yielding not only helps in 
understanding the physics of glasses~\cite{wisitsorasak2017dynamical,jin2018stability}, but also
guides the design 
of amorphous materials for industrial applications~\cite{Telford_2004, Greer_2013}.

Yielding in amorphous solids is broadly classified into two classes:
ductile and brittle yielding. Ductile
materials accumulate plastic activity before yielding and show a smooth
crossover from elastic-to-plastic regimes.  
Many soft glassy materials, such as foams~\cite{Lauridsen_2002},
emulsions~\cite{Mason_1996}, and colloidal
glasses~\cite{schall2007structural} belong to this class. 
In brittle materials, yielding is characterized by 
a sharp stress overshoot accompanied  
by strain localization, which triggers material failure at large deformation. 
Metallic glasses~\cite{hufnagel2016deformation,Greer_2013},
window glasses, and concretes~\cite{Sim_2014} are typical brittle
materials. 

Brittleness is not necessarily an intrinsic material property. For example, brittleness can gradually increase by changing the cooling rate to prepare the system~\cite{Utz_2000,Fan_2017, Shen_2007, Ashwin_2013,rodney2011modeling,ketkaew2018mechanical} or by physical aging~\cite{Volynskii_2007, Rottler_2005, Hasan_1993, Liu_2010,
Varnik_2004,Moorcroft_2011}. 
In athermal conditions, this evolution was recently described as a phase transition
and confirmed by atomic glass simulations under uniform shear~\cite{Ozawa_2018}. Evidence was provided that a critical point 
separates ductile and brittle behaviors, demonstrating the
importance of glass stability to understand yielding. 

In many studies of the yielding transition, a large amplitude
oscillatory shear (LAOS) protocol is used instead of a uniform shear
flow. This is useful to relate macroscopic properties with microscopic
trajectories of the constituting atoms~\cite{Hyun_2011}. LAOS is also
the first step toward understanding the response of amorphous solids
under dynamical loading conditions~\cite{Forquin_2017}, relevant to the
rheology of soft materials like emulsions and
colloids~\cite{Knowlton_2014, denisov2015sharp}, and the
mechanism of fatigue failure in metallic
glasses~\cite{wang2013real,sha2015cyclic}. Previous numerical work under
oscillatory shear found that yielding 
in athermal conditions
is associated with a microscopic
irreversible transition at the particle
level~\cite{Kawasaki_2016,Regev_2013,Regev_2015,Fiocco_2013}. 
It was also shown that small strain amplitude oscillatory shear can increase glass
stability~\cite{Fiocco_2013, Leishangthem_2017, Parmar_2019, Das_2019,Schinasi-Lemberg2020pre}.  
Such ``mechanical annealing''~\cite{Priezjev_2018_moldyn} was
proposed as a route distinct from thermal annealing to
prepare stable
glasses~\cite{Das_2018}. When the strain amplitude is increased above
yielding, the system cannot find a stable energy state, and the plastic
activity associated with energy dissipation is observed at each
cycle~\cite{Fiocco_2013,Kawasaki_2016, Leishangthem_2017}.   
Large-scale simulations~\cite{Fiocco_2013, Parmar_2019,
Priezjev_2018_yt} and a mesoscopic model~\cite{Radhakrishnan_2016,
Radhakrishnan_2018} in this regime showed that macroscopic shear bands
can form at long times,
via a mechanism proposed to be similar to shear band formation under uniform shear at large deformation~\cite{shi2005strain,
Shrivastav_2016_hd,Shrivastav_2016_yt}. 
However, the role of glass preparation on the nature of yielding under oscillatory shear has not been tested over a broad range of glass stabilities.
Therefore, the relative role of thermal and mechanical annealings on mechanical properties remains unclear~\cite{Kawasaki_2016, Parmar_2019, Fiocco_2013, Leishangthem_2017, Das_2019, Das_2018, Priezjev_2018_yt, Priezjev_2018_moldyn}.

Our goal is to unify the yielding transition of thermally and mechanically annealed systems and to study how yielding changes when glass preparation is varied over a broad enough range to mimic the physics of both colloidal and metallic glasses under oscillatory shear. We use atomistic computer simulations of a simple glass former over a wide range of preparation temperatures~\cite{Ninarello_2017}. We carefully disentangle the relative roles of mechanical and thermal annealings, and the various physical regimes accessible in the (strain amplitude, preparation temperature) phase diagram, regarding both the nature of the yielding transition and the distinct mechanisms of shear band formation in different glassy materials. 


We simulate a three-dimensional size polydisperse system~\cite{Ninarello_2017} with a pairwise soft
potential given by 
  \begin{equation}\label{eq:energy}
    U(r_{ij}) = \epsilon_0 \Bigg[ \bigg(\frac{d_{ij}}{r_{ij}}\bigg)^{12} + c_0 + c_1 \bigg(\frac{r_{ij}}{d_{ij}}\bigg)^2 + c_2 \bigg(\frac{r_{ij}}{d_{ij}}\bigg)^4 \Bigg],
  \end{equation}
where $r_{ij} =|\bold{r}_i - \bold{r}_j|$ 
is the distance between particles $i$ and $j$, and  
$d_{ij} \equiv (d_i + d_j)(1 - 0.2|d_i - d_j|)/2$ controls the nonadditive interaction between particles of diameters $d_i$ and $d_j$. The cutoff distance of the potential is set to $r^{(ij)}_\text{cut} = 1.25 d_{ij}$, 
and the constants $c_0$, $c_1$, and $c_2$ are chosen so that the first and second derivatives of the potential vanish at the cutoff. We, respectively, use the average diameter $\overline{d}$ and energy scale $\epsilon_0$ as our length and energy units. To generate amorphous solids with a broad range of stabilities, we first produce equilibrium configurations of the supercooled liquid over a wide range of temperatures $T_{\text{init}}$, using the swap Monte Carlo method~\cite{Ninarello_2017}. We study the range $T_\text{init} \in [0.062, 5]$ and work at fixed  
number density $\rho = 1.02$. These equilibrium configurations are then instantly quenched to zero temperature (using the \texttt{FIRE}
algorithm~\cite{Bitzek_2006}) to form the initial glass configurations
to be sheared. The model and physical properties of these glasses were
documented in Refs.~\cite{Ozawa_2018,
Ninarello_2017,Coslovich_2018,Wang_2019}. In particular, the energy of
the inherent structures, $E_{\rm IS}$, is almost constant at high $T_{\text{init}}$, and starts decreasing as $T_{\text{init}}$ drops below the onset temperature $T_\text{onset}\approx 0.18$. Several physical properties change qualitatively when $T_\text{init}$ decreases below the mode-coupling crossover temperature, $T_\text{MCT} \approx 0.108$~\cite{Wang_2019}. 
The estimated experimental glass temperature is 
$T_\text{g}\approx 0.072$.

We deform these zero-temperature amorphous solids using a simple oscillatory shear with a strain amplitude $\gamma_0$ at constant volume. 
We employ both the athermal quasistatic (AQS) shear protocol~\cite{maloney2006amorphous} and the finite (but small) strain rate protocol. A homogeneous shear strain is realized using Lees-Edwards boundary conditions~\cite{Lees_1972}. In the AQS protocol, the system is strained at each step by a small incremental strain $|\Delta \gamma| = 2 \times 10^{-4}$, after which the configuration is again relaxed into the nearest energy minimum. In the finite 
strain rate protocol, the system is
driven by an overdamped dynamical equation~\cite{Kawasaki_2016} 
\begin{equation}\label{eq:eom}
 \xi \bigg[
\frac{{d} \textbf{r}_i}{{d} t} 
- \dot{\gamma}(t) \textbf{r}_{i}(t)\cdot \hat{\textbf{y}}\hat{\textbf{x}} 
\bigg] 
+ \sum_{j \neq i} \frac{\partial U(r_{ij})}{\partial \textbf{r}_{i}} = \textbf{0},
  \end{equation}
where $\hat{\bf x}$ and $\hat{\bf y}$ are the
unit vectors in the $x$ and $y$ directions, $\textbf{r}_i$ is the 
position of particle $i$, and $\xi$ is the friction coefficient that sets the time unit $\overline{d} \xi / \epsilon_0$.
The applied strain is $\gamma (t) = \gamma_0 \sin (2\pi t / T_\text{cyc})$, with an oscillation period $T_\text{cyc}$. Most reported results are for the 
finite strain rate protocol with a small strain rate $\dot{\gamma}_0
\equiv 2 \pi \gamma_0 / T_\text{cyc} = 6.2832 \times 10^{-4}$,  
and total number of particles $N = 12000$. 
In the Supplemental Material, 
we show that this finite strain rate is small enough to reproduce
results obtained from the AQS limit \cite{SI}.
Where needed, we also discuss the
results of AQS simulations and larger systems $N = 48000$.  
A complete system size dependence is presented in the Supplemental Material
\cite{SI}.
For each data point, we average over at least 50 cycles for at least
three different initial conditions after the system reaches a steady state.  
Unless specified otherwise, error
bars in each figure represent the standard deviation of the measurements.


\begin{figure}
  \includegraphics[width=8.5cm]{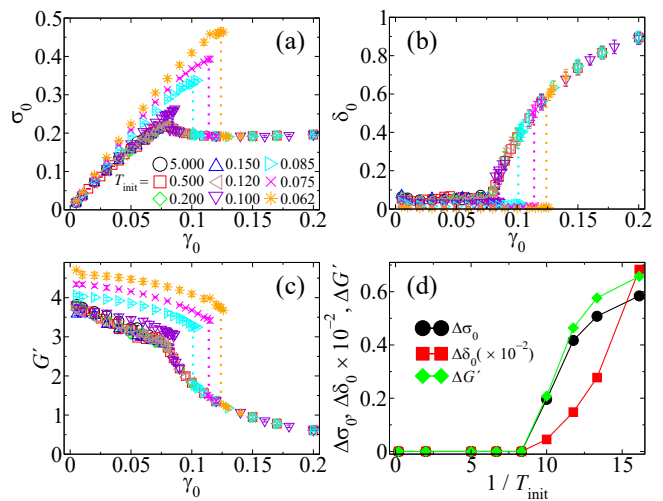} 
    \caption {
       (a)--(c) Various rheological quantities as a function of the strain 
       amplitude $\gamma_0$ for different
       $T_\text{init}$'s [see the legend shown in (a)]. 
       (a) Stress amplitude $\sigma_0$. 
       (b) Phase lag $\delta_0$.
       (c) Storage modulus $G^\prime$. 
Dashed vertical lines in these figures represent the discontinuous jumps of these quantities.
 (d) The magnitude of the discontinuities of these quantities at yielding 
        as a function of $1/T_\text{init}$ vanishes at $T_\text{c} \approx 0.1$. 
}   
  \label{f:response}
\end{figure}

We first present the behavior of macroscopic rheological quantities in
steady state after many cycles. We calculate the linear description of the nonlinear stress-strain relation~\cite{Hyun_2011} by fitting the time-dependent shear stress $\sigma(t)$ (see the Supplemental Material for the definition \cite{SI}) to $\sigma_0 \sin (2\pi t /T_\text{cyc} + \delta_0)$, where $\sigma_0$ is the stress amplitude and $\delta_0$ is the phase lag. We then  obtain the storage modulus as $G' = \sigma_0 \cos (\delta_0) / \gamma_0$. 
The $\gamma_0$ dependence of these quantities is shown in
Figs.~\ref{f:response}(a-c) for several preparation temperatures
$T_{\rm init}$. 
Data for the less annealed systems, $0.12 \leq T_{\rm init} \leq 5$,
whose stabilities would correspond to typical soft materials~\cite{Ozawa_2018}, essentially collapse on curves that are continuous with a cusp at the same yield strain amplitude $\gamma_\text{Y} \approx 0.08$ irrespective of $T_{\rm init}$;  
In Fig.~\ref{f:response}(a) $\sigma_0$ displays a smooth overshoot at $\gamma_\text{Y}$, consistent with Refs.~\cite{Fiocco_2013, Leishangthem_2017, Kawasaki_2016};
$\delta_0$ starts to increase sharply in Fig.~\ref{f:response}(b), but 
almost continuously at $\gamma_\text{Y}$. In Fig.~\ref{f:response}(c) $G^\prime$ decreases with a kink at $\gamma_\text{Y}$, as also shown in Ref.~\cite{Kawasaki_2016}. 
A qualitatively distinct behavior is observed when 
$T_\text{init}$ is below a critical value $T_\text{c} \approx 0.1$.
The data at small $\gamma_0$ no longer collapse and are distinct for each
 $T_{\rm init}$.
One observes that more stable glasses have larger
 shear and storage moduli, and are less dissipative than poorly annealed
 ones. At the  yielding transition point, all data
display a sharp discontinuity,
which depends on the glass stability. 
To our knowledge, such large discontinuities were not
 observed before in computer simulations under oscillatory shear,
 because it was impossible to produce stable enough glasses~\cite{Fiocco_2013, Leishangthem_2017, Parmar_2019, Das_2019,Schinasi-Lemberg2020pre}. 
Beyond yielding, all curves merge again, as the plastic activity drives the
 glass away from its stable initial conditions, and memory is eventually
 lost~\cite{Fiocco_2013, Leishangthem_2017}. 
Interestingly, the number of cycles required to reach the steady state remains finite below $T_\text{c}$ in the vicinity of $\gamma_\text{Y}$, 
whereas it diverges for poorly annealed
systems~\cite{Kawasaki_2016,Regev_2013,Fiocco_2013} (see the Supplemental Material~\cite{SI}).
In Fig.~\ref{f:response}(d), we show the magnitude of the jumps of the
 various rheological quantities
at $\gamma_\text{Y}$ as a function of 
$T_{\rm init}$ 
(See the Supplemental Material for the precise definitions~\cite{SI}). 
This representation eloquently demonstrates that the nature of
 yielding changes qualitatively at the critical value of 
$T_\text{c}\approx 0.1$. This value is close to the critical temperature discussed in the context of uniform shear~\cite{Ozawa_2018}, but more precise measurements and a finite size scaling analysis would be needed to establish a more precise connection. 

\begin{figure}
  \includegraphics[width=8.5cm]{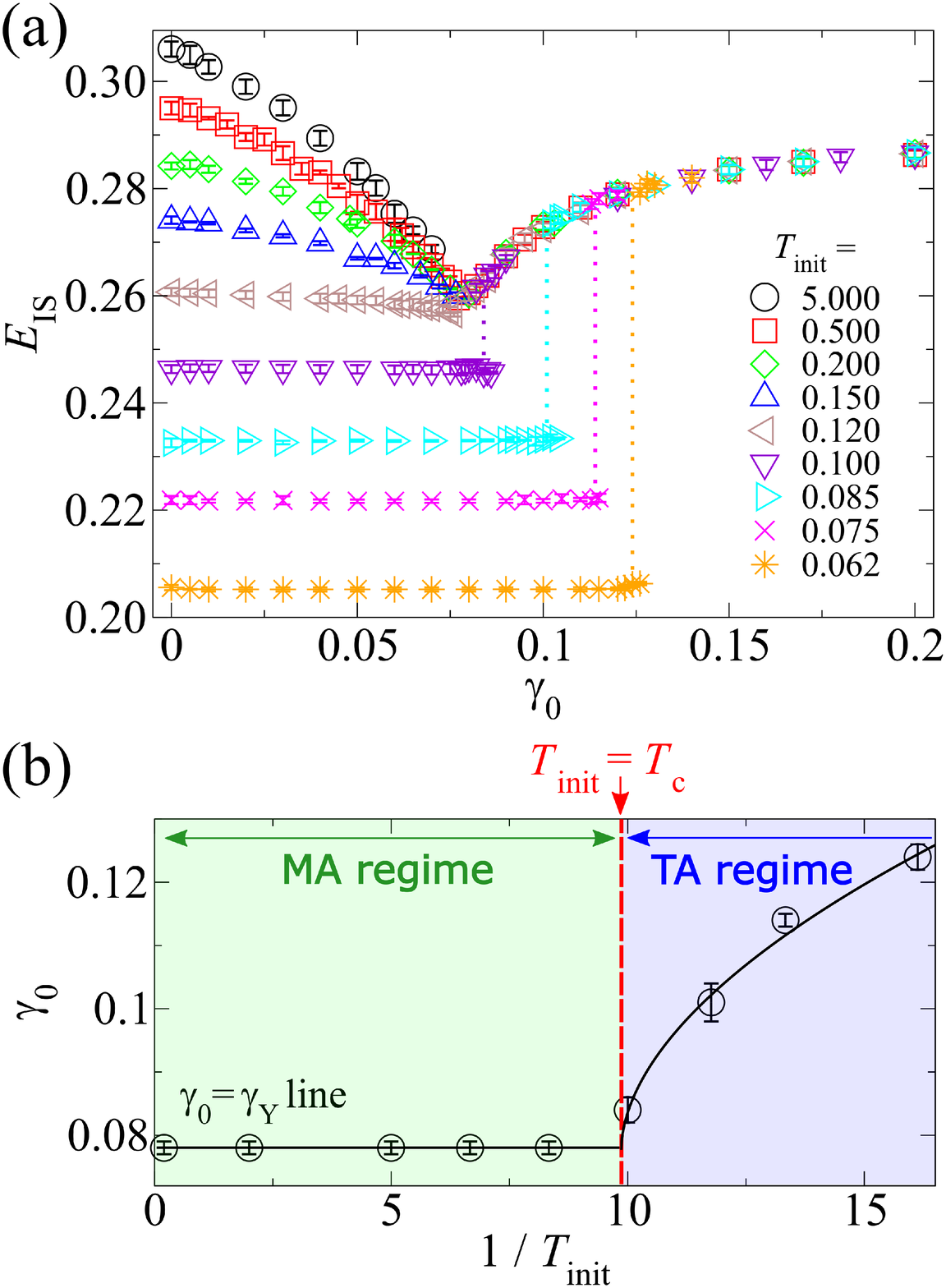}
    \caption {
      (a) Steady-state inherent structure energy $E_\text{IS}$
      as a function of the strain amplitude $\gamma_0$  
      for several $T_\text{init}$. The dashed vertical lines represent the discontinuous jumps.
      (b) Nonequilibrium phase diagram in the 
$(\gamma_0,T_{\rm init}^{-1})$ plane. The critical temperature $T_\text{c}$ (red
 long dashed line) divides the phase space into the mechanical annealing
 (MA) and thermal annealing (TA) regimes. 
The black solid line separates the elastic and yielded regions (see the text).
 } 
  \label{f:phase}
\end{figure}

To analyze the relative role of thermal and mechanical annealings, we
study the dependence of the steady-state inherent structure energy, $E_{\rm IS}$, 
on $T_{\rm init}$ and $\gamma_0$, see Fig.~\ref{f:phase}(a). 
We again find two qualitatively distinct behaviors, 
depending on the value of $T_{\rm init}$ before yielding. 
For $\gamma_0=0$, $E_{\rm IS}$ decreases with decreasing $T_{\rm init}$, 
reflecting the increasing stability of the initial conditions by thermal annealing~\cite{Wang_2019}. 
For $T_{\rm init} > T_\text{c}$, $E_{\rm IS}$ is a decreasing function of
$\gamma_0$, which confirms that poorly annealed glasses can access
deeper energy states when submitted to oscillatory shear of modest
amplitude. This is mechanical annealing (MA), as reported in previous
works~\cite{Das_2018,Priezjev_2018_moldyn}. 
By contrast, for $T_{\rm init} < T_\text{c}$ \cite{MAcomment}, mechanical annealing is not observed. 
This implies that thermal annealing (TA) is so efficient for these glasses that mechanical annealing is no longer able to drive them toward lower energy states. The recent results of Ref.~\cite{Schinasi-Lemberg2020pre} can be interpreted as representative of glasses aged near or slightly below $T_c$. 

Driving poorly annealed glasses above yielding produces instead higher
energy configurations, and the yielding transition for those materials
appears therefore as a cusp in $E_{\rm IS}$, whereas it again appears as
a sharp discontinuity when $T_{\rm init} < T_\text{c}$, which emerges at $T_c$.
We conclude that mechanical annealing operates
for small $\gamma_0$ and high enough $T_{\rm init}$, 
but becomes inefficient when the effect of
thermal annealing increases at low $T_{\rm init}$. 
The emergence of the critical temperature $T_\text{c}$ can be physically understood as a
direct consequence of this competition. 
It corresponds to the temperature below which mechanical annealing is no
longer useful.
Since $T_\text{c}$ corresponds to a sharp change in the nature of yielding, we see no conceptual reason to relate it to the mode-coupling temperature, which describes a smooth physical crossover in finite dimensions and equilibrium conditions~\cite{srikanth}. Physically, lowering $T_c$ decreases the quenched disorder and presumably the density of weak regions in the initial glass configurations~\cite{Ozawa_2018}.

\begin{figure}[h]
  \includegraphics[width=8.5cm]{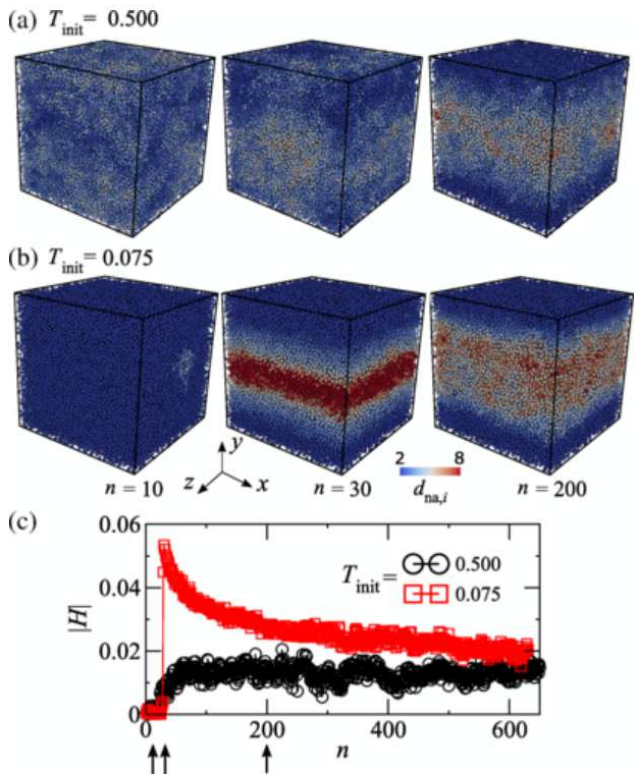} 
    \caption {
      The snapshots show a color map of the one-cycle nonaffine
      deformation $d_{\text{na},i}$ for (a) $T_\text{init} = 0.5$ and (b) $0.075$ after  
      $n = 10$, $30$, and $200$ cycles (from top to bottom).
      (c) Evolution of the shear band order parameter $|H|$.  
      The arrows indicate the times shown in (a, b). Yielding in (a) is accompanied by the slow emergence of a diffusive shear band, whereas in (b) a sharp shear band is suddenly formed when the material fails macroscopically.
      Data obtained using the AQS protocol and $N=48000$ with 
$\gamma_0 = 0.11$.}
  \label{f:band}
\end{figure}

These findings can be summarized in the nonequilibrium phase diagram
in the $(\gamma_0,T_{\rm init}^{-1})$ plane shown in Fig.~\ref{f:phase}(b). 
The yield strain amplitude $\gamma_\text{Y}$ separates the
``elastic'' region at low strain amplitude where the plastic activity is
suppressed in steady state, from the ``yielded''
region where plastic events occur during each cycle at large
strain amplitude. $\gamma_\text{Y}$ is constant in the regime dominated
by mechanical annealing (MA regime) for 
$T_{\rm init} >T_\text{c}$~\cite{Leishangthem_2017}, 
but it increases with decreasing $T_{\rm init}$ in the regime dominated
by thermal annealing (TA regime). 
The value of $T_\text{c}$ can be determined by a power-law fit of
$\gamma_\text{Y}$~\cite{SI}, which gives $T_\text{c} \approx 0.101$. 
Although the nature of yielding changes dramatically with $T_{\rm init}$, it is
accompanied for all stabilities by a similar discontinuous irreversible
transition at the particle scale (see, for example, Fig.~S9 in the Supplemental Material \cite{SI}).   

The existence of two distinct regimes for yielding under oscillatory
shear also influences the nature of shear banding above yielding.
Recent works on the poorly annealed glasses in the AQS limit and small
strain rate suggested that persistent shear bands may emerge under slow oscillatory
shear for large systems at long times~\cite{Parmar_2019, Fiocco_2013, Priezjev_2018_yt}. 
We confirm that a persistent shear band forms in our simulations at long times in similar conditions,
but it does not appear when the strain rate is large or $\gamma_0$ is
away from $\gamma_\text{Y}$~\cite{SI}. 
In Fig.~\ref{f:band}(a), we illustrate the slow emergence of a
shear band after many cycles using AQS simulations and  larger system
size, $N=48000$, starting from a poorly annealed glass with 
$T_{\rm init}=0.5$. In steady shear, the stress-strain rate relation is monotonic and no shear band is formed~\cite{singh2020brittle} and thus the observed shear band purely originates from the oscillatory nature of the  shearing.
A color map of the one-cycle nonaffine deformation $d_{\text{na},i}$
for each particle $i$ \cite{Schreck2013pre} 
is used to visualize strain localization after many cycles. 
To quantify the gradual formation of the
shear band, we introduce an order parameter $|H|$, which quantifies
the extent of spatial inhomogeneities of 
$d_{\text{na},i}$. See the Supplemental Material for the precise definition of $d_{\text{na},i}$ and $|H|$ \cite{SI}.
In Fig.~\ref{f:band}(c), we report the
corresponding slow increase of $|H|$. 
The finite value of $|H|$ at steady state
implies
the presence of a persistent shear band.

Shear band formation is very different for $T_{\rm init} < T_\text{c}$, 
as demonstrated
in Fig.~\ref{f:band}(b). 
In the first few cycles, a very small amount of localized plastic activity is observed. 
This is followed by a sudden, macroscopic failure of the material
accompanied by the instantaneous formation of a shear band at $n = 29$
cycles. 
The shear band is formed within a single cycle, which is very
different from the gradual emergence reported for $T_{\rm init} > T_\text{c}$ at long times~\cite{Parmar_2019, Fiocco_2013, Priezjev_2018_yt}.
The observed behavior is reminiscent of fatigue failure in metallic
glasses under cyclic loading in the sense that, after some loading
cycles, the system fails abruptly~\cite{zhang2004fatigue}. 
The time dependence of $|H|$ in
Fig.~\ref{f:band}(c) confirms the sudden formation of the band, which
gradually broadens as the number of cycles increases. 
At very long times, the memory of the initial condition is lost and
both types of systems display similar dynamics.  
 

In this work, we systematically analyzed the mechanical properties under
oscillatory shear of amorphous solids prepared at different depth
in the energy landscape and the interplay between
thermal and mechanical annealings. We have identified a critical temperature $T_\text{c}$ separating two distinct regimes for yielding. For systems prepared at $T_\text{init} >T_\text{c}$ in the MA regime (corresponding to typical soft materials), mechanical
annealing can drive the system toward lower energy states at small strain amplitude, but memory is gradually lost at large strain amplitude. In the TA regime, $T_\text{init} < T_\text{c}$ (corresponding to typical metallic glasses), no mechanical annealing occurs at small strain amplitudes, 
and yielding corresponds to a macroscopic failure of the material
accompanied by a discontinuous jump of macroscopic rheological
quantities and energy, associated with the sudden appearance of a
shear band. Therefore, the MA and TA regimes are separated by a
ductile-to-brittle transition.  
The present work dealing with oscillatory shear extends qualitatively
similar findings obtained for uniform shear in AQS conditions~\cite{Ozawa_2018} to time-dependent flows.  

We observed distinct mechanisms of shear band formation in the MA and TA
regimes. A shear band forms gradually at long times in the MA regime,
becoming persistent in the steady state after yielding.
In the TA regime, however, an abrupt emergence of a shear band in a single cycle is found. In particular, before shear band formation, a small number of localized precursor events 
are observed (see Fig.~\ref{f:band}(b)),
which would initiate a macroscopic
shear band~\cite{priezjev2019shear}. Future work should address in more
detail the distinct mechanisms for shear banding in these two
regimes. 
Some preliminary results on this aspect are shown in the Supplemental Material \cite{SI}.

We investigated the important role of glass stability on the yielding under oscillatory shear relevant for various experimental situations from the rheology of soft materials to fatigue failure in metallic glasses. Our work elucidates the interplay between mechanical and thermal annealing processes, providing useful information for material design in industrial applications~\cite{sun2016thermomechanical}, as well as theoretical descriptions of amorphous materials under cyclic deformation~\cite{perchikov2014variable,Radhakrishnan_2016, Radhakrishnan_2018}.

\acknowledgments
We thank Srikanth Sastry for discussions. 
We also thank the anonymous referee for bringing our attention to Ref.~\cite{Schinasi-Lemberg2020pre}.
This research is supported by Japan Society for the Promotion of Science (JSPS) KAKENHI
(Nos.~16H06018,   
	16H04034,   
	18H01188,    
	19K03767,    
and  19H01812) and a grant from
the Simons Foundation (No.~454933, L.~B.).

\bibliography{References}

\end{document}